\documentclass[aip,pre,amsmath,amssymb,reprint,longbibliography]{revtex4-1}
\usepackage{hyperref}
\usepackage{graphicx}
\usepackage{dcolumn} % align table columns on decimal point
\usepackage{color}
\usepackage{bm} % bold math

\newcommand{\der}[2]{\frac{d#1}{d#2}}
\definecolor{apsblue}{rgb}{0.18,0.19,0.57}
\definecolor{darkblue}{rgb}{0.2,0.1,0.5}
\definecolor{darkgreen}{rgb}{0.1,0.6,.1}
\definecolor{darkred}{rgb}{0.7,0.0,.1}

\hypersetup{
 colorlinks,
 citecolor=apsblue,
 urlcolor=apsblue,
 linkcolor=apsblue,
 pdfpagemode={UseNone},
 pdfauthor={},
 pdftitle={},
}

\graphicspath{{figures/}:{./}}

\begin{document}

\title{Dynamic and thermodynamic crossover scenarios in the Kob-Andersen mixture: Insights from multi-CPU and multi-GPU simulations}

\author{Daniele Coslovich}

\author{Misaki Ozawa}

\author{Walter Kob}

\affiliation{Laboratoire Charles Coulomb, Universit\'e de Montpellier, CNRS, Montpellier, France}

\begin{abstract}
The physical behavior of glass-forming liquids presents complex features of
both dynamic and thermodynamic nature.  Some studies indicate the presence
of thermodynamic anomalies and of crossovers in the dynamic properties,
but their origin and degree of universality is difficult to assess.
Moreover, conventional simulations are barely able to cover the range of
temperatures at which these crossovers usually occur. To address these
issues, we simulate the Kob-Andersen Lennard-Jones mixture using efficient protocols based on
multi-CPU and multi-GPU parallel tempering.
Our setup enables us to probe the thermodynamics and dynamics of the liquid at equilibrium well below the critical temperature of mode-coupling theory, $T_{\rm MCT} = 0.435$.
We find that below
$T=0.4$ the analysis is hampered by partial crystallization of the
metastable liquid, which nucleates extended regions populated by large particles
arranged in an fcc structure.  By filtering out crystalline samples,
we reveal that the specific heat grows in a regular manner down to
$T=0.38$. Possible thermodynamic anomalies suggested by previous
studies can thus occur only in a region of the phase diagram where
the system is highly metastable. Using the equilibrium configurations
obtained from the parallel tempering simulations, we perform molecular
dynamics and Monte Carlo simulations to probe the equilibrium dynamics
down to $T=0.4$.  A temperature-derivative analysis of
the relaxation time and diffusion data allows us to assess different
dynamic scenarios around $T_{\rm MCT}$.  Hints of a dynamic crossover
come from analysis of the four-point dynamic susceptibility.  Finally,
we discuss possible future numerical strategies to clarify the nature
of crossover phenomena in glass-forming liquids.

\end{abstract}

\maketitle

\section{Introduction}
\label{sec-2}

If a liquid is cooled quickly enough, it will bypass crystallization and form an amorphous solid called glass. 
Such a glass transition is directly related to the rapid increase of the structural relaxation
time upon cooling, whose temperature-dependence 
is given by the Arrhenius law for strong glass-formers and
is super-Arrhenius for fragile glass-formers~\cite{angell1995formation}.
Despite being purely kinetic in nature, the glass transition is accompanied by
a change in the thermodynamic properties of the system at the glass
transition temperature $T_g$. For instance, the specific heat shows
a mild increase upon cooling and displays a drop at $T_g$ due to the freezing
of the configurational degrees of freedom. Although no phase transition
occurs at $T_g$, it has been suggested that the slowing down of the
dynamics and the resulting glass formation might be driven by a hidden
phase transition~\cite{binder2011glassy,berthier2011theoretical}.

While most glass-forming liquids follow the trends outlined above,
there are notable exceptions. This is the case for instance of
liquids with strong, directional interactions, such as silica,
silicon, and water. These systems display a thermodynamic
anomaly in the form of a local maximum of the specific heat
at a temperature $T^*>T_g$~\cite{angell_glass-formers_2008},
a behavior which has also been observed in simulation studies of
silica~\cite{scheidler_frequency-dependent_2001,saika-voivod_free_2004},
supercooled water~\cite{saito_frequency_2013}, and other simple
models~\cite{moreno_energy_2005,xu_monatomic_2009,gutierrez2015static,ozawa2016tuning}.
Concomitantly, also the dynamic behavior of these liquids changes
around $T^*$, crossing over from a super-Arrhenius to an Arrhenius
temperature dependence (``fragile-to-strong'' crossover). The
origin of these thermodynamic and dynamic anomalies has often
been attributed to the presence, or proximity, of a liquid-liquid
transition or its Widom-line, located at temperatures higher
than $T_g$~\cite{angell_glass-formers_2008}.  Interestingly,
this fragile-to-strong crossover is not limited to 
systems with directional interactions~\cite{mallamace_transport_2010}.
Recent experimental~\cite{zhang_fragilestrong_2010-1,zhou2015structural}
and simulational~\cite{lad_signatures_2012} studies have
demonstrated that also several metallic glass-formers exhibit a
notable fragile-to-strong crossover.  Several interpretations have
been proposed to explain the crossover in these systems, including
the presence of thermodynamic anomalies such as a liquid-liquid
transition~\cite{lad_signatures_2012,wei_liquidliquid_2013,stolpe2016structural},
anomalous crystallization~\cite{yang2014anomalous}, or the evolution of
structural medium range order~\cite{zhou2015structural}.

The dynamics of some molecular and polymeric liquids show even a
different kind of crossover at a temperature $T_{\rm D}$, typically
located about 15\%-25\% above $T_g$.  This dynamic crossover thus
occurs when the relaxation times are around $10^{-8}-10^{-6}$~s
and is subtle, but it can be revealed by temperature-derivative
analysis of high-quality dynamic measurements in molecular
glass-formers~\cite{stickel_dynamics_1995,martinez-garcia_new_2012,novikov_qualitative_2015}.
At $T_{\rm D}$ the increase of relaxation times crosses over
from super-Arrhenius to a milder temperature dependence and
tend to become Arrhenius at very low temperature. While
some authors have considered systems with such a crossover
as marginal cases~\cite{elmatad_corresponding_2009},
an alternative point of view suggests that this behavior
may fairly general~\cite{mallamace_transport_2010},
even though the location of the crossover is highly
system-dependent~\cite{martinez-garcia_new_2012,novikov_qualitative_2015}.

The physical origin of this dynamic crossover is not completely
clear and there are diverging viewpoints on this. Some authors
have pointed out the closeness of $T_{\rm D}$ with the temperature
at which a power law fit predicts a divergence of the relaxation time
data~\cite{casalini_dynamic_2003,casalini_viscosity_2004,mallamace_transport_2010}.
The dynamic crossover should then be identified with the critical
temperature of mode-coupling theory (MCT)~\cite{gotze2008complex}. Others
have attributed this crossover to an upper limit of the activation
energy~\cite{elmatad_corresponding_2009,novikov_qualitative_2015},
which may saturate at an arbitrarily low temperature. While these two
interpretations are completely different in nature, they are difficult
to disentangle in practice.  We emphasize that the dynamic crossover
scenario is a priori unrelated from one that accompanies liquid-liquid
transitions, in that the former does not involve the thermodynamics and
is purely dynamic in origin.

Also computer simulations have been used to probe the existence of
anomalies in the dynamic and thermodynamic properties of glass-forming
liquids. The results of these studies suggest that both the dynamic
correlation length scales~\cite{kob_non-monotonic_2012} and the
dynamic finite size effects~\cite{berthier_finite-size_2012}
may show a crossover compatible with several theoretical
predictions~\cite{biroli2012random,rizzo2015qualitative,rizzo2016dynamical},
although these results do not seem to hold universally. On the other
hand, the dynamic range accessible in conventional numerical studies is
limited to only 4-5 decades in time. Thus, some of these results might be
affected by insufficient equilibration.

In the present study, we focus on a simple glass-former, proposed
by Kob and Andersen (KA)~\cite{kob_testing_1995}, that has so far
been fairly robust against crystallization and hence has been
used in many investigations of the glass-transition. Previous
studies of this system have given evidence for the presence of
a thermodynamic anomaly in the form of a peak in the specific
heat~\cite{flenner_hybrid_2006} and a possible fragile-to-strong
crossover~\cite{ashwin_low-temperature_2003,doliwa_energy_2003},
although these results might have been affected by finite size and
finite sampling effects. In the present work we employ an
optimized simulation setup, which exploits the parallel tempering
method~\cite{hukushima_exchange_1996} and state-of-art molecular
dynamics code running on graphics processing units (GPU), to extend
the temperature range in which thermodynamic and dynamic measurements
can be done at equilibrium. We find that, in contrast to previous
reports~\cite{flenner_hybrid_2006}, the thermodynamics of the liquid is
regular. At lower temperatures, simulations are hindered by
crystallization, which involves structures of fcc symmetry formed by
large particle.  Finally, we discuss possible ways and setups to detect
numerically the presence of dynamic anomalies in model glass-formers.

\section{Model and methods}
\label{sec:methods}
\subsection{Model parameters}
The system we consider is a binary mixture in which both species have
the same mass $m$.  The particles interact via a Lennard-Jones potential
given by
\begin{equation}\label{eqn:lj}
u_{\alpha\beta}(r) = 4\epsilon_{\alpha\beta}\left[
  {\left( \frac{\sigma_{\alpha\beta}}{r} \right)}^{12} - {\left(
    \frac{\sigma_{\alpha\beta}}{r} \right)}^6 \right] \quad ,
\end{equation}
where $\alpha, \beta \in \{\rm A,B\}$ are species indices.  The value
of the parameters $\sigma_{\alpha\beta}$ and $\epsilon_{\alpha\beta}$
are given in Ref.~\cite{kob_testing_1995}.  The units of length and
energy are set by the parameters $\sigma_{\rm AA}=1$ and $\epsilon_{\rm
AA}=1$, respectively.  The potentials are cut and shifted at a
distance $2.5\sigma_{\alpha\beta}$.  We simulate systems composed
by $N$ particles in a cubic box of side $L$ with periodic boundary
conditions and a number density given by $\rho=N/V=1.1998$. Note that
even small differences in density can quantitatively affect the
thermodynamic and dynamic observables.  For example, in the original
paper~\cite{kob_testing_1995} a different density $\rho=1.204$ was
used, which leads to slight differences in statics and dynamics when
sufficiently high quality data are available.  The system size ranges
from $N=300$ to $3600$ for parallel tempering simulations (see below).
Additional dynamic data have been obtained for a much bigger sample
($N=100000$) using the LAMMPS simulation package~\cite{LAMMPS,LAMMPS-code}.

\subsection{Simulation protocols}
Our simulations implement the parallel tempering (PT)
algorithm~\cite{hukushima_exchange_1996,yamamoto_replica-exchange_2000},
in which $M$ replicas of the system of interest perform independently
simulations at temperatures $\{T_i\}$ with the potential energies $\{
U_i \}$.  At regular intervals, exchanges are attempted between pairs
of replicas at neighboring states and the temperatures are exchanged
with probability 
\begin{equation} p = {\rm min} \left\{ 1, \exp
\left[     (U_i - U_j)\left( \beta_i - \beta_j \right) \right] \right\},
\end{equation}
where $\beta_i=1/ k_{\rm B} T_i$ (with $k_{\rm B}=1$
in this study), to ensure detailed balance.  Since our simulations
extend to a temperature range that remained so far largely unexplored,
we used different simulation protocols and software to check and validate
our results.  Namely, we used two independent implementations of the PT
algorithm, each one relying on a different molecular dynamics code
to carry out the simulation.
Note that, for a given system size, we used the same sets of temperatures for all the PT protocols, namely
0.3730, 0.3810, 0.3901, 0.4003, 0.4115, 0.4238, 0.4374, 0.4525, 0.4692, 0.4877, 0.5082, 0.5307 for $N=1200$ and
0.4000, 0.4060, 0.4130, 0.4210, 0.4301, 0.4403, 0.4515, 0.4638, 0.4768, 0.4906 for $N=3600$.

In the first implementation, named \textbf{PT-1 protocol}, we perform
multi-CPU parallel tempering simulations with an in-house molecular
dynamics code.  The MD simulations are performed in the $NVT$ ensemble
using the Nose-Hoover thermostat~\cite{frenkel2001understanding} with
a time step $\delta t=0.004$ and a thermostat relaxation time $0.4 = 100 \delta
t$.  We used $M=12-14$ replicas
depending on system size. The code is parallelized using MPI to handle
communications between replicas, which attempt to exchange their state,
i.e., the temperature of the associated thermostat, every $50000$ MD steps.

The second implementation, named \textbf{PT-2 protocol}, relies on a
multi-GPU parallel tempering code~\cite{atooms-pt} that uses the RUMD package~\cite{rumd}
as a simulation backend. This multi-GPU code was implemented in python 
building on the \texttt{atooms} framework~\cite{atooms}.
Multiple replicas are simulated on individual GPUs and communication between GPUs are handled at high-level via the \texttt{mpi4py} package~\cite{mpipy}.
We ran the multi-GPU simulations on a dedicated cluster of inexpensive gaming cards (GTX-980 and even GTX-750Ti).
The thermostat is again of the Nose-Hoover type,
the time step is $\delta t=0.004$ and the thermostat relaxation time
$\tau_T = 0.2 = 50 \delta t$.  We used the same exchange intervals and
number of replicas as in PT-1.  We checked that increasing the interval
between exchanges did not change our results.

Finally, we performed additional PT simulations (\textbf{PT-3 protocol})
to extend our thermodynamic measurements to even lower temperatures than
the ones attained by protocols PT-1 and PT-2. The PT-3 simulations are
performed using the multi-GPU code described above starting from uncorrelated configurations
obtained using the PT-2 protocol at a temperature $T=0.4$. 
These efficient multi-GPU simulations enabled us to carefully
measure the waiting time dependence of the results so as to assess
equilibration issues, see Sec.~\ref{sec:thermo}.

\begin{figure}
  \centering
  \includegraphics[width=1\linewidth]{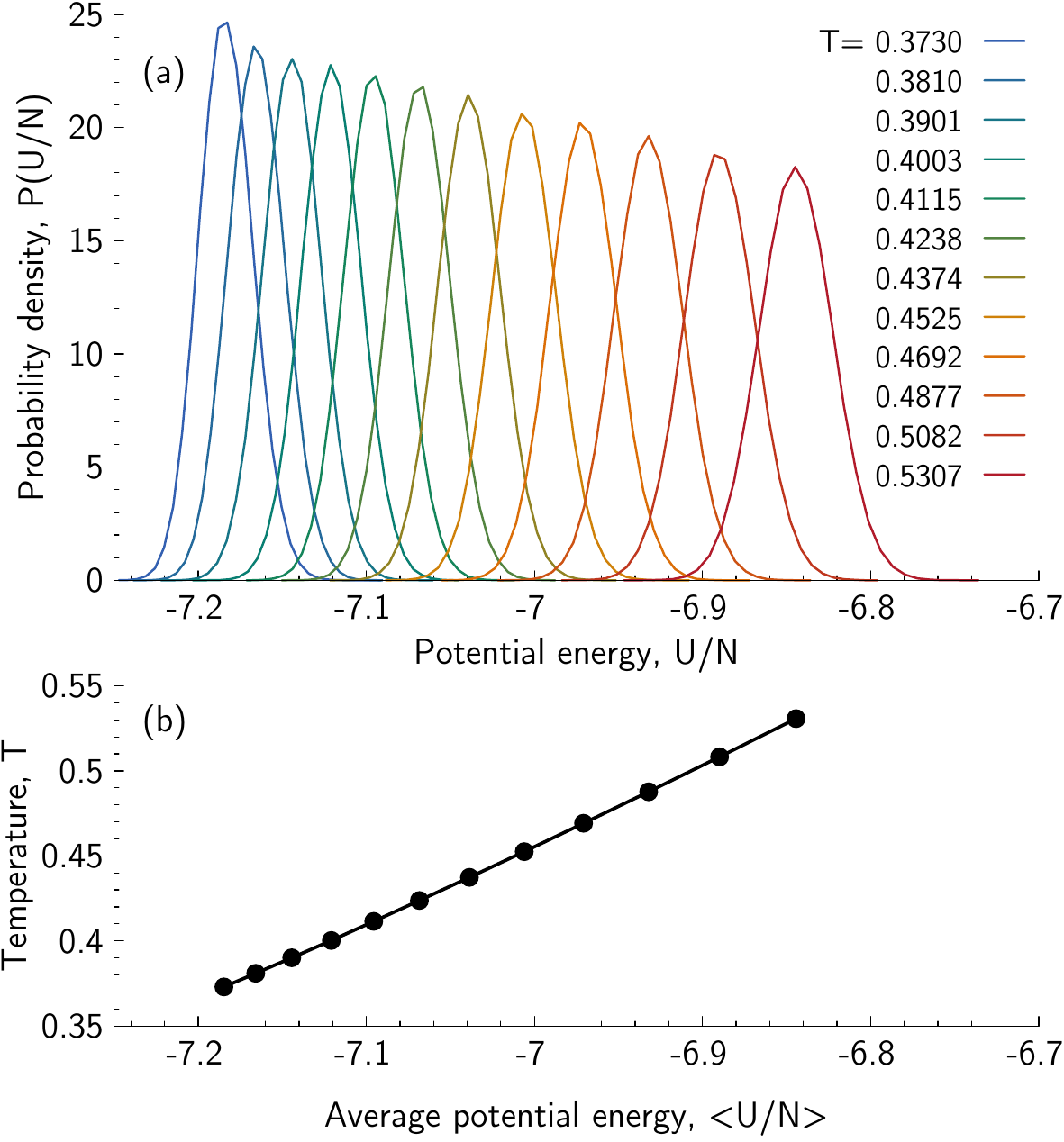}
  \caption{\label{fig:energy}(a) Distribution of the potential energy per particle $P(U/N)$ for $N=1200$ obtained from the parallel tempering protocol PT-3. Runs with a substantial fraction of crystalline configurations were discarded from the analysis. (b) Average potential energies per particle for each of the studied temperatures.}
\end{figure}

From these PT simulations, we evaluated the specific heat per particle $c_V$ from the fluctuations of the potential energy $U$,
\begin{equation}
c_V =  \frac{1}{N T^2} \left( \langle  U^2 \rangle - \langle  U \rangle^2 \right) + \frac{3}{2} ,
\end{equation}
and from the temperature derivative of the average potential energy,
\begin{equation}
c_V = \frac{1}{N} \frac{\partial \langle U \rangle}{\partial T} + \frac{3}{2} ,
\end{equation}
where $\langle (\cdots)\rangle$ is the thermal average.
The two expressions yield identical results provided the averages are carried out over the equilibrium measure.
In a simulation, the agreement between the estimates of $c_V$ obtained through the two methods is often taken as a test of equilibration.
We note that, in practice, only very accurate measurements can reveal discrepancies between the two methods.
The distributions $P(U/N)$ and the $T$-dependence of $\langle U \rangle$ obtained from well-equilibrated simulations (PT-3 protocol) are shown in Fig.~\ref{fig:energy}.

Strictly speaking, parallel tempering simulations only give access to thermodynamic and static properties.
However, it is possible to perform extended dynamic measurements by starting ``regular'' simulations from configurations sampled during the PT simulations at a given temperature.
Here, again, we follow two distinct protocols to corroborate our results.

In the \textbf{MC protocol} we performed normal Monte Carlo simulations using an in-house code and starting from uncorrelated configurations obtained from protocol PT-1.
The MC simulations are carried out in the $NVT$ ensemble using simple displacement moves~\cite{berthier2007monte},
in which we attempt to displace a randomly selected particles over a cube of side $0.15$.
We used $10-30$ independent configurations depending on temperature.
The length of the simulations at the lowest temperature ($T=0.4$) is $10^9$ Monte Carlo steps.
In the following, the time unit for the MC protocol is given by one MC sweep, i.e., $N$ attempted displacement moves.

In the \textbf{MD protocol} we performed molecular dynamics simulations using the RUMD package starting from uncorrelated  configurations obtained from protocol PT-1 and PT-2.
The MD simulations are carried out in the $NVT$ ensemble using the Nose-Hoover thermostat, the time step is $\delta t=0.004$ and the thermostat relaxation time $\tau_T = 0.2 = 50 \delta t$.
For the $N=1200$ samples, we used $128$ independent configurations down to $T=0.4$ from protocol PT-2.
For $T=0.39$, we used only $30$ configurations, namely the final configurations of the PT-2 runs.
For the $N=3600$ samples, we used $20$ independent configurations from protocol PT-1.
The duration of each of the MD simulations ranged from $4.2\times 10^6$ steps (at $T=0.51$) to $2.1\times 10^9$ steps (at $T=0.39$), thus
each run was about $10-20$ times longer than the typical structural relaxation time $\tau_\alpha$ (see below for its definition).
In total, for each temperature, our simulations cover over about 2500 structural relaxation times.
This high-quality statistics enables us to perform a temperature-derivative analysis of the dynamic data, see Sec.~\ref{sec:dynamics}.

For both MC and MD protocols, we checked that the initial
configurations were uncorrelated from one another by measuring their
mutual self overlaps~\cite{donati2002theory}
\begin{equation}
  Q_{\rm s} =  \frac{1}{N} \sum_i \Theta(a- |{\bf r}_i^{\alpha} - {\bf r}_i^{\beta}|),
\end{equation}
where $\alpha$ and $\beta$ denote two configurations, and their mutual collective overlaps
\begin{equation}
  Q_{\rm c} =  \frac{1}{N} \sum_{i,j} \Theta(a - |{\bf r}_i^{\alpha} - {\bf r}_j^{\beta}| ).
\end{equation}
A sensible choice of parameter $a$ is a fraction of the typical
interparticle distance. We chose $a=0.3$. We found that both $Q_{\rm
s}$ and $Q_{\rm c}$ are close to the values expected for uncorrelated pairs
of configurations, i.e., $O(1/N)$ and $\frac{4}{3}\pi a^3\rho$, respectively.

\begin{figure}
  \centering
  \includegraphics[width=1\linewidth]{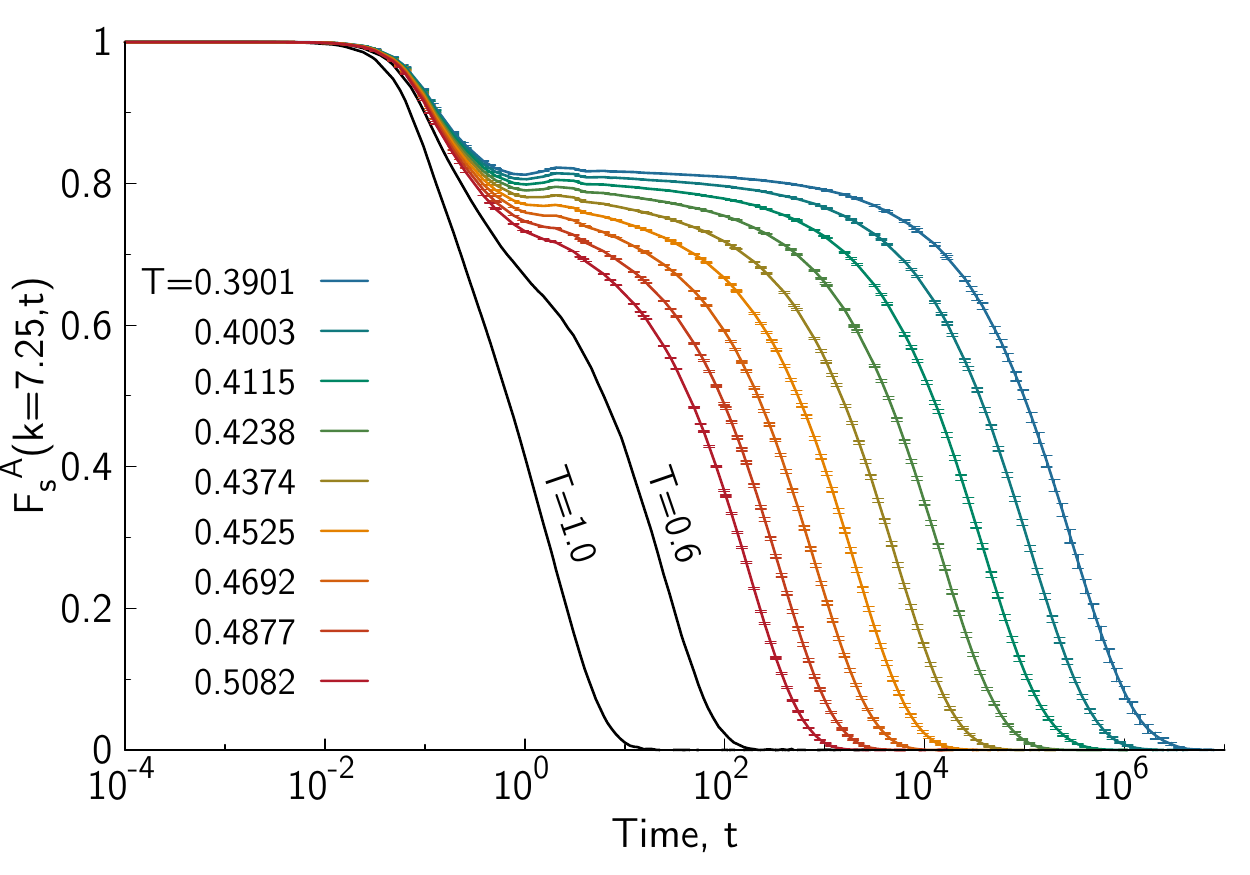}
  \caption{\label{fig:fskt}
    Self part of the intermediate scattering functions $F_s^A(k,t)$ obtained from the MD
    protocol for $N=1200$ particles. Errors bars are one standard deviation on the mean, calculated over about 128 runs.}
\end{figure}

From MD and MC simulations we extract the self part of the intermediate scattering functions
\begin{equation}
  F_s^A(k,t)= \langle  f_s^A(k,t) \rangle =  \left\langle 
    \frac{1}{N_A}\sum_j e^{-i\textbf{k}\cdot[\textbf{r}_j(t)-\textbf{r}_j(0)]} \right\rangle,
\end{equation}
where the sum runs over the particles of type $A$. We choose
a wave-vector $k=7.25$, close to the first peak of the structure
factor~\cite{kob_testing_1995}.  The corresponding structural relaxation
time $\tau_\alpha$ is defined as usual as $F_s^A(k,\tau_\alpha)=1/e$.
In Fig.~\ref{fig:fskt}, we show the dynamic data obtained from the MD protocol.

\subsection{Crystalline order detection}

\begin{figure}[!h]
  \centering
  \includegraphics[width=\linewidth]{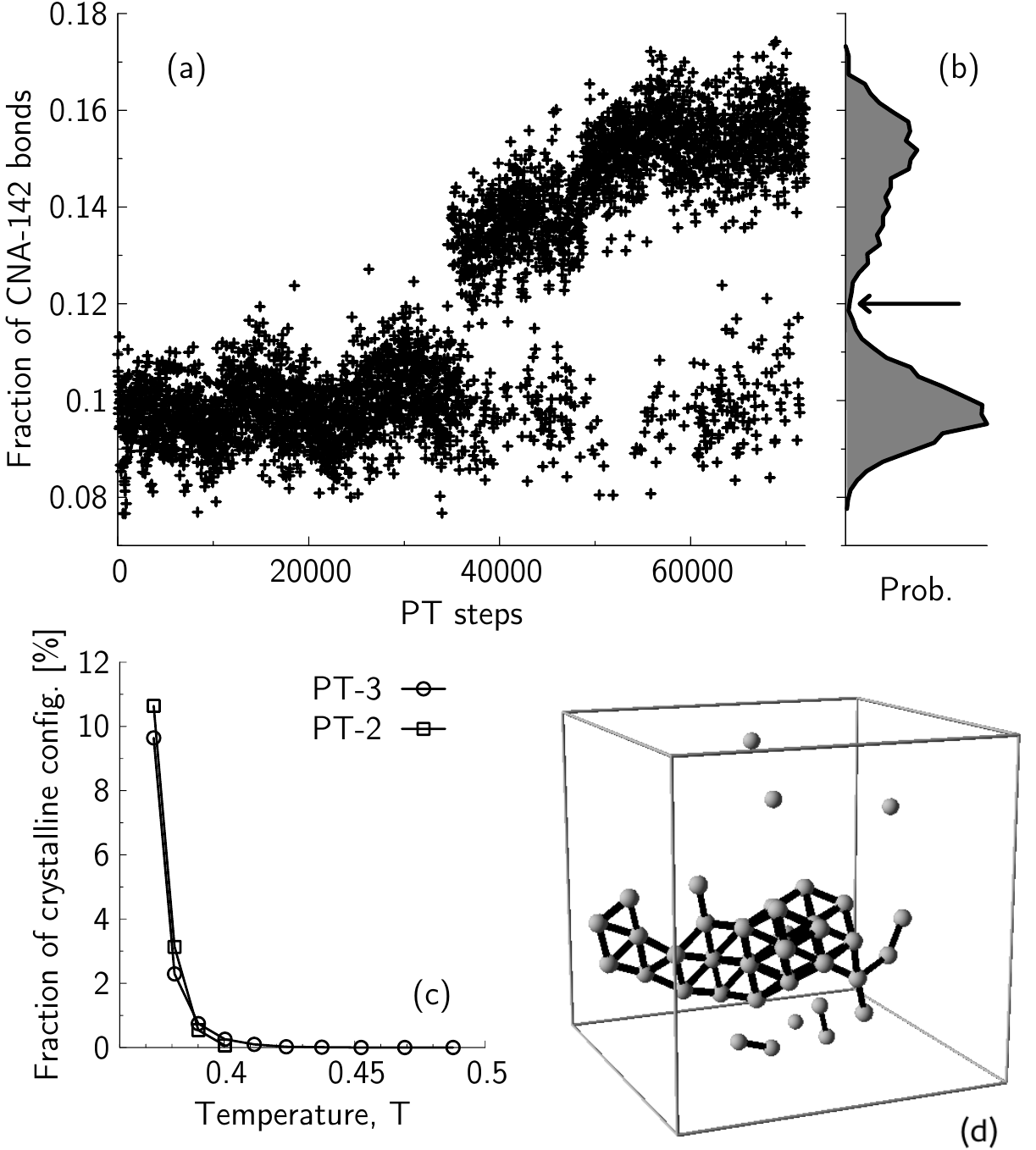}
  \caption{\label{fig:xtal}
    Detection of crystalline configurations. 
    (a) Crystallization event during a PT-2 simulation for $N=1200$ particles. The fraction of CNA-142 bonds, $f_{142}$, is shown as a function of PT steps (1 PT step=50000 MD steps). (b) The probability density calculated during the run shows a bimodal distribution with a sharp minimum around $f_{142}\approx 12\%$. (c) Percentage of crystalline configurations, detected using a 12\% threshold on $f_{142}$, as a function of temperature. Below $T=0.4$, both PT-2 and PT-3 protocols have a large fraction of crystalline configurations. PT-2 data were not analyzed above $T=0.4$. (d) Connected component of a cluster formed by particles surrounded by $A$-particles only during a crystallization event. }
\end{figure}

The study of glass-forming liquids is often hampered by crystallization
and the very relation between glassy behavior and crystallization remains
a matter of debate~\cite{kawasaki2010formation,tanaka_bond_2012}.  The KA
mixture, which is a simple model of a metallic glass-former, has been
extensively used as a model to study the glass transition because of its
stability against crystallization. Until very recently, the note
added in the proofs of Ref.~\onlinecite{toxvaerd_stability_2009} was, to the best of our knowledge, the only
report of crystallization of this model by direct simulation, achieved
through runs of about $3.7\times 10^7$ time units ($7.4\times 10^9$ steps)
at $T=0.40$.  
At this temperature, however, the nucleation time is still
much larger than the structural relaxation time $\tau_\alpha$ ($\sim 10^5$
time units), and therefore MD / MC simulations of the metastable liquid
can be carried out safely. In this work, however, we were able to equilibrate the mixture at
even lower temperatures. Below $T=0.4$, crystallization
events become increasingly frequent, as also demonstrated by a very recent simulation study~\cite{Ingebrigtsen_Dyre_Schrøder_Royall_2018}. 
Within the studied range of system sizes, the smaller the system, the stronger the tendency to crystallization.

\begin{figure*}[!ht]
  \centering
  \begin{tabular}{ccc}
  \includegraphics[width=0.33\linewidth]{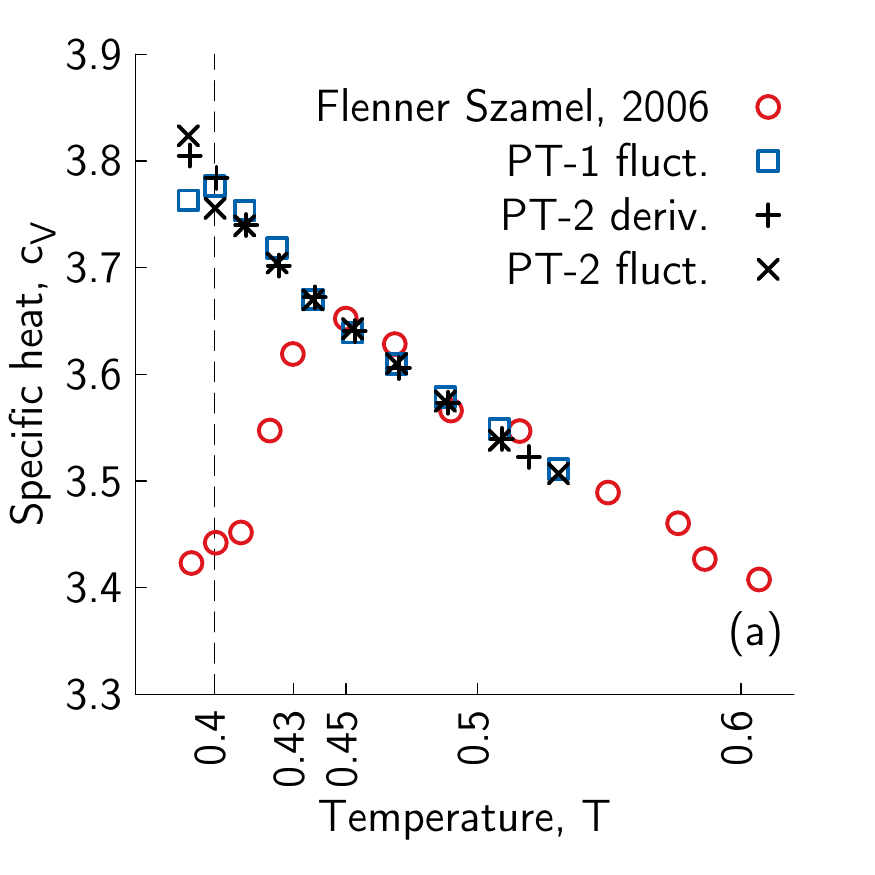} & 
  \includegraphics[width=0.33\linewidth]{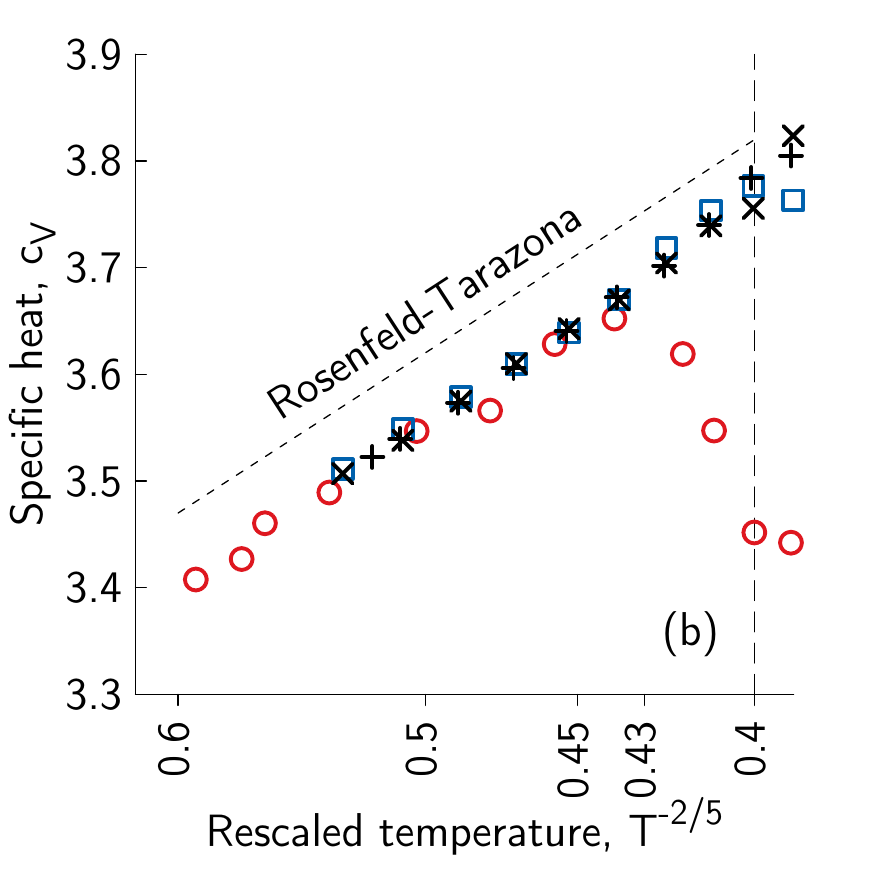} &
  \includegraphics[width=0.33\linewidth]{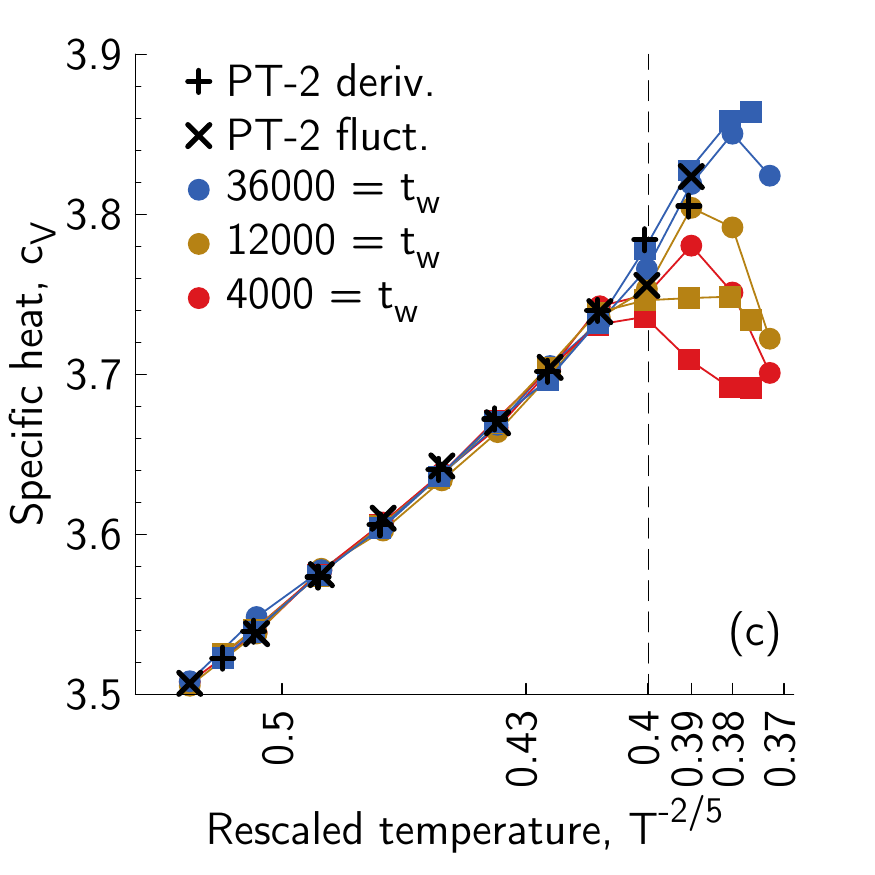} \
  \end{tabular}
  \caption{\label{fig:cv} (a) Specific heat from protocols PT-1 and PT-2 for $N=1200$ particles in the regime $T \gtrsim 0.4$ where crystallization is negligible. Results from fluctuation and derivative expressions of the specific heat are shown as indicated in the legend. The temperature of the data from Ref.~\onlinecite{flenner_hybrid_2006} (circles) is divided by the density scaling factor $(1.204/1.2)^\gamma$, with $\gamma=5.0$ to correct for the small density mismatch. (b) Rosenfeld-Tarazona scaling of the same specific heat $c_V$ data as in panel (a). (c) Specific heat from protocol PT-3 for different waiting times $t_w$ expressed in unit of PT steps. Results from fluctuation and derivative expressions of the specific heat correspond to circles and squares, respectively. Configurations with $f_{\rm 142} > 0.12$ are discarded from the calculations.}
\end{figure*}

As a first indicator of crystallization events in our simulations,
we monitored the evolution of the inherent structure (IS) energy
as a function of time~\cite{stillinger_hidden_1982}.  However,
the IS energy may also display large but reversible fluctuations,
unrelated to crystal nucleation.  We thus studied two additional
order parameters that allow us to disentangle ``amorphous''
and crystalline fluctuations.  The first one relies on the so-called
common neighbor analysis (CNA)~\cite{honeycutt_molecular_1987}.  In this
approach, the bonds formed by neighboring particles are classified
according to the number of shared neighbors.  It has been shown that
the fraction $f_{\rm 142}$ of bonds of type 142, see e.g. Ref.~\onlinecite{hedges_dynamic_2009}, allows one to detect crystallization
in \textit{biased} simulations of the KA mixture.  We found that this
approach allows one to detect crystallization in the bulk mixture as well.
An example of a crystallization event is depicted in Fig.~\ref{fig:xtal}.
Even though the nature of the fluctuation is not always clear-cut,
we found that a threshold on $f_{142}$ is
an effective criterion to filter out crystalline configurations.
Note that since we run several replicas at a time, only a few of them
may be affected by crystallization. When this occurs they typically
remain ``stuck'' in the lowest portion of temperature space.  As in
Ref.~\onlinecite{hedges_dynamic_2009}, we used a threshold of 12\% CNA-142 bonds.
A large crystalline cluster detected in our simulations is shown in
Fig.~\ref{fig:xtal}.  The crystal nucleus is formed by fcc pockets of
$A$ particles, which implies compositional fluctuations that deplete
$B$ particles.  To detect it, we introduced an even simpler order
parameter, namely the concentration of cages formed by $A$ particles
only~\cite{hedges_dynamic_2009}.  In particular, we evaluated the
connected component~\cite{newman_networks:_2010} of clusters formed by
pure-$A$ cages.  We found that in typical crystalline samples, the size
of these connected clusters is about a few hundred particles.  Finally,
in Fig.\ref{fig:xtal}(c) we show the percentage of crystalline configurations
in the simulations of protocol PT-2 and PT-3 for $N=1200$ particles.
We see that they increase markedly below $T=0.4$.  To deal with data
analysis in this delicate regime while retaining the maximum possible
amount of statistics, we filtered our $c_V$ measurements by discarding
individual configurations whose CNA-142 concentration was higher than
0.12, see Sec.~\ref{sec:thermo}.  On the other hand, above $T=0.4$ the fraction of samples above
the crystalline threshold is negligible and we
rarely encountered problematic runs. These runs were simply
discarded all together.

\section{Results}
\subsection{Thermodynamics}\label{sec:thermo}

The specific heat $c_V$ is a sensitive, although not unambiguous,
indicator of thermodynamic changes in dense liquids.  For instance, a
sudden drop in $c_V$ as a function of the control parameters may be due to
a phase transition, or more generally to a change in the topography of
the underlying energy landscape, but might as well indicate incomplete
equilibration, as is the case at the laboratory glass transition observed
upon cooling.  Previous work on the KA mixture showed the presence of a
peak in $c_V$ at some temperature close to the mode-coupling temperature
$T_{\rm MCT}=0.435$~\cite{kob_testing_1995}.  In this section, we show that the thermodynamics of
the KA mixture is regular, i.e., it shows no anomaly, at least down to $T=0.39$ and that the peak
found in Ref.~\onlinecite{flenner_hybrid_2006} is due to partial equilibration.
Below $T=0.4$, equilibration becomes hard on current simulation time
scales and analysis is further hampered by crystallization events.
Finally, we discuss the possible presence of a thermodynamic anomaly in this 
highly metastable portion of the phase diagram.

In Fig.~\ref{fig:cv}(a) we show specific heat measurements from parallel
tempering simulations using protocols PT-1 and PT-2.  We also include
results from a previous study using PT simulations, which reported a
peak in $c_V$ around $T_{\rm MCT}$~\cite{flenner_hybrid_2006}.  From our data, we conclude that no
anomaly is observed in the specific heat, which increases monotonically
down to $T=0.39$.  The length of our simulations is typically one order
of magnitude longer than those of Ref.~\onlinecite{flenner_hybrid_2006}.
In retrospect, our work warns that tests such as histogram
reweighting or the consistency of fluctuation and derivative
expressions of response functions, such as $c_V$, are not sufficient to ensure equilibration of supercooled liquids,
see Ref.~\onlinecite{odriozola_berthier_2011} for a more detailed discussion of this issue.
Also, our results imply that one
needs at least $\sim 10^9$ MD time steps to equilibrate the KA mixture
below $T_{\rm MCT}$.

The empirical model of Rosenfeld-Tarazona~\cite{Rosenfeld_Tarazona_1998} proposes the following
functional form for the potential energy $U=aT^{3/5}+b$, which yields
$c_V\sim T^{-2/5}$.  In Fig.~\ref{fig:cv}(b) we draw the specific heat
data from protocols PT-1 and PT-2 as a function of $T^{-2/5}$, which
linearizes the Rosenfeld-Tarazona law.  A similar representation
yields an excellent data collapse for simple liquids in both normal
and moderately supercooled regime~\cite{Ingebrigtsen_Veldhorst_Schroder_Dyre_2013}.  We find that this functional form
provides a very good description of the data but a slight upward bending is observed around $T=0.4$, 
suggesting the presence of additional fluctuations
not accounted for by this simple liquid-state model.
This is confirmed by the analysis of the third moment of the potential energy distribution. We found that the skewness remains constant at high temperature and starts to increase slightly around $T=0.4$ (not shown), even when crystalline configurations are removed from the analysis. This behavior may be thus attributed to crystallization precursors or to enhanced fluctuations of the locally favored structure of the system~\cite{coslovich_locally_2011,Turci_Royall_Speck_2017}.

We found that the measurements using PT-1 and PT-2 start to deviate
significantly from each other below $T=0.39$, Moreover, in this
temperature regime, the liquid has an increased tendency to crystallize,
see Sec.~\ref{sec:methods}.  In Fig.~\ref{fig:cv}(c) we show measurements of
$c_V$ from protocol PT-3, in which crystalline samples are removed on a per-configuration basis and
equilibration is assessed directly, i.e., by measuring the waiting
time dependence of $c_V$. Specifically, we start parallel tempering
simulations from previously equilibrated configurations at $T=0.4$ and perform
averages over restricted portions of the trajectories as follows
\begin{equation}
  \langle A \rangle_{t_w}  = \frac{1}{t_p-t_x} \int_{t_w}^{t_w+t_p} \mathrm{d} t A(t) W(t),
\end{equation}
where $t_w$ is the waiting time and $W(t)=\Theta(0.12 - f_{\rm 142})$ is a windowing function
that removes from the averages samples identified as crystalline
(see Sec.~\ref{sec:methods}) and $t_x = \int_{t_w}^{t_w+t_p} \mathrm{d}t
(1-W(t))$. The production time $t_p$ was kept fixed to 36000 PT steps, independent of waiting time.

In Fig.~\ref{fig:cv}(c) we show the results obtained by calculating the
specific heat starting from time $t_w$ with increasing values of
$t_w$.  We emphasize that these specific heat measurements cover a
temperature regime that has never been probed before at equilibrium.
For temperatures higher than 0.4, the results of the PT-3 protocol
show a consistent, i.e., $t_w$-independent, growth of $c_V$, thus
corroborating our previous analysis.  Below $T=0.4$, the specific heat
measurements display a more marked waiting-time dependence but fluctuation
and derivative formulas converge for sufficiently long waiting times and
reveal a monotonic increase of $c_V$ down to $T =0.38$.  Only at
the lowest temperature, the specific heat $c_V$ measured from energy
fluctuations shows a local maximum.  The dependence of this peak as a
function of waiting time suggests that this feature may be actually due
to lack of equilibration.

\begin{figure*}[!ht]
  \centering
  \begin{tabular}{ccc}
  \includegraphics[width=0.33\linewidth]{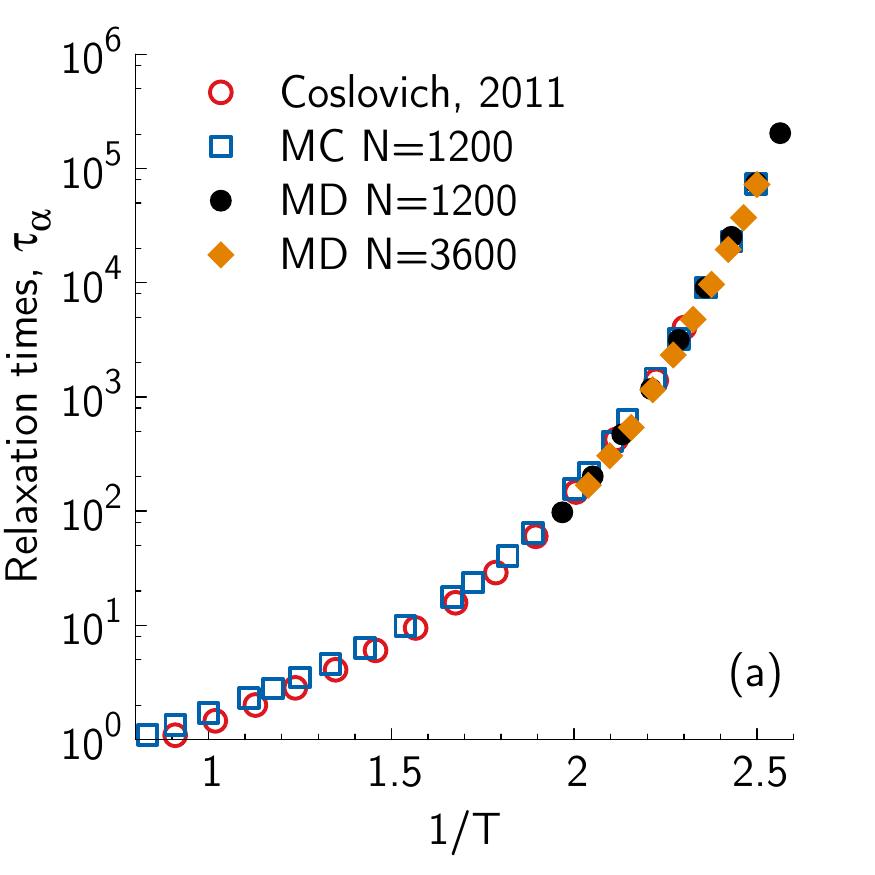} &
  \includegraphics[width=0.33\linewidth]{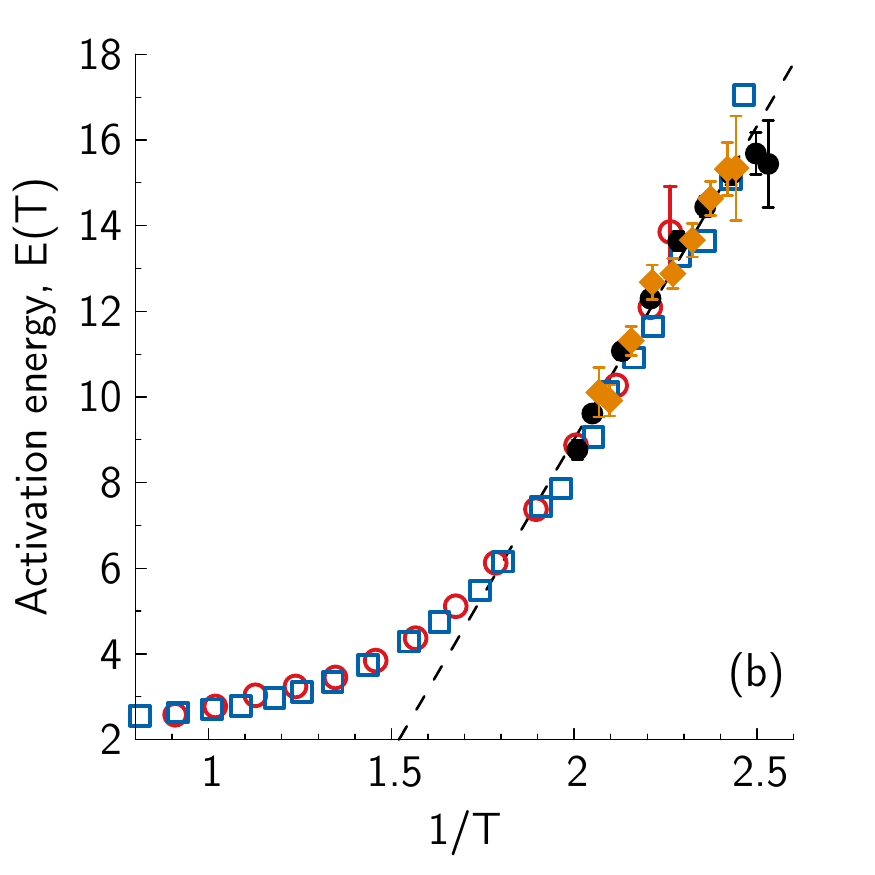} &
  \includegraphics[width=0.33\linewidth]{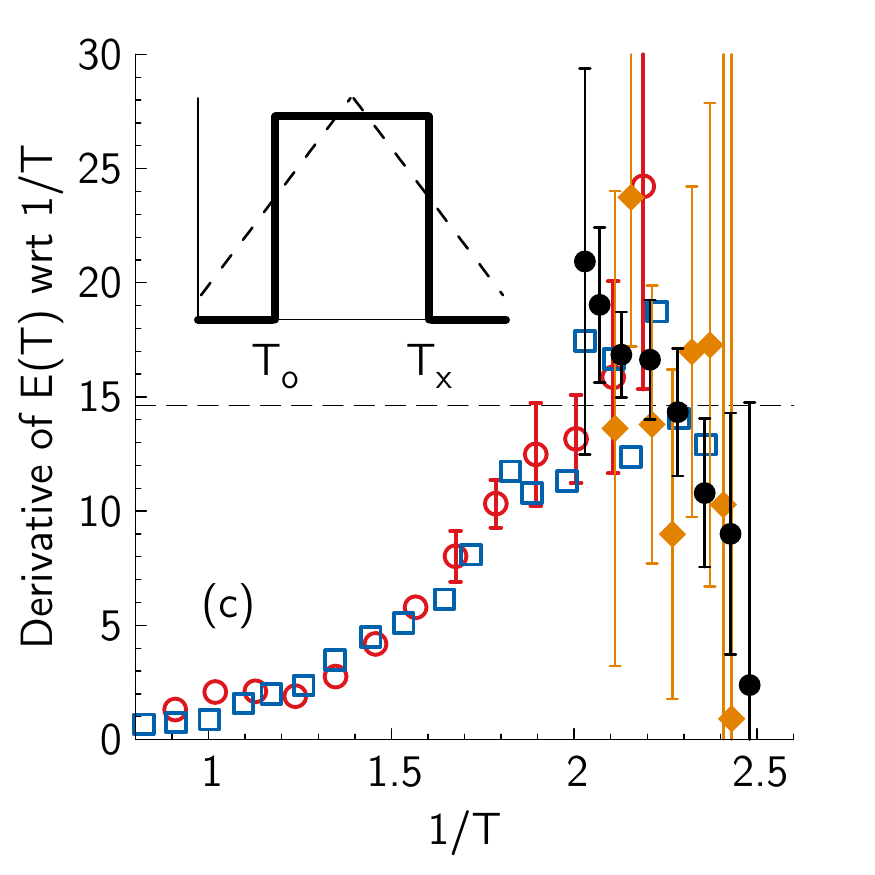} \\
  \end{tabular}
  \caption{\label{fig:tau}Temperature-derivative analysis of relaxation times. The relaxation times obtained from MC simulations have been scaled by $5\times 10^2$ MC steps. Error bars on MD data represent one standard deviation on the mean and are only shown when larger than the symbol size. (a) Relaxation times $\tau_\alpha$ as a function of $1/T$ for various protocols and system sizes. (b) Activation energy $E(T)$ for the same set of data as panel (a). The dashed line indicates a linear fit to Eq.~\eqref{eqn:paraE} in the range $T\le 0.52$ with $J=2.7$ and $T_0=0.72$. (c) Derivative of $E(T)$ with respect to $1/T$. The horizontal line corresponds to $2J^2$. The inset shows a schematic representation of two possible scenarios for this derivative: the solid line depicts the behavior expected from facilitation models, while the dashed line is the qualitative behavior described in Ref.~\onlinecite{novikov_qualitative_2015}.}
\end{figure*}

Recent simulations based on trajectory path sampling~\cite{Turci_Royall_Speck_2017} suggest the
existence of a liquid-liquid transition in the low temperature part of
the phase diagram of the KA mixture. Our data narrow down the temperature range over which this hypothetical transition may occur and rule out thermodynamic anomalies for $T>0.37$. These results do not exclude, however, the scenarios discussed
in Ref.~\onlinecite{Turci_Royall_Speck_2017}. Given the strong tendency to crystallize
below $T=0.4$, however, our results show that a thermodynamic anomaly,
if present at all in the KA mixture, is hidden in a highly metastable
portion of the phase diagram. In practice, it will be very difficult to
detect it in large enough samples. Therefore, we think that future work
in this context should focus on more robust models of glass-formers.

\subsection{Dynamics}
\label{sec:dynamics}

The parallel tempering algorithm allows one to accelerate the sampling of
static and thermodynamic observables, but it does not provide by itself
useful information on the dynamics.  However, it is possible to carry out
dynamic measurements by using uncorrelated configurations extracted from
PT simulations as starting points of conventional molecular dynamics
or Monte Carlo simulations.  This approach can be parallelized in a trivial way by performing independent simulations and allows one to extend the
accessible dynamic range by short-circuiting equilibration issues.
In this section, we implement these ideas
following using the MD and MC protocols
described in Sec.~\ref{sec:methods} and test various dynamic crossover
scenarios below the MCT temperature.

Figure~\ref{fig:tau}(a)
shows an Arrhenius representation of the relaxation times obtained using the
MD and MC protocols for $1200$ and $3600$ particles. 
We note that equilibrium sampling is ensured by the fact that both MD and MD simulations start from previously equilibrated configurations down to $T=0.39$.
The time scale of Monte Carlo simulations has been scaled to
match the relaxation times of MD at low temperature. We also include
dynamic data from Ref.~\onlinecite{coslovich_locally_2011}, which agree
with the new set of simulations over the common temperature
range. Very similar behavior is found also for B-type particles
(not shown). Thanks to the hybrid protocols employed in the present work, the accessible dynamic
range has increased by almost two orders of magnitude compared to the
conventional molecular dynamics simulations of Ref.~\onlinecite{coslovich_locally_2011} and we can
comfortably study the dynamics below $T_{\rm MCT}$. 
The data suggest that the temperature dependence of the relaxation times gets milder, i.e., more
Arrhenius-like, at the lowest temperatures. However, it is notoriously
difficult to draw firm conclusions based on analysis of the relaxation
time alone, which has led to a number of controversies~\cite{mallamace_transport_2010}.

Temperature-derivative analysis of the dynamic data provides a
most stringent test of analytic expressions for $\tau_\alpha(T)$
and is particularly well-suited to reveal the presence of a dynamic
crossover~\cite{stickel_dynamics_1995,martinez-garcia_new_2012,novikov_qualitative_2015}.  Even though this approach
requires very accurate data, it has the advantage of being parameter
free and requires no data fitting.  So far, it has mostly been applied
to high-quality dielectric relaxation measurements.  To the best of our
knowledge, the only numerical study to conduct this analysis across
the MCT temperature is Ref.~\onlinecite{kob_nonlinear_2014} for a mixtures
of harmonic spheres.  Here, we push this kind of analysis even further
by combining the trivial parallelism of our MD and MC protocols
and the efficiency of the RUMD simulation package on small system
sizes~\cite{rumd}. 

The central quantity in our analysis is the apparent activation energy
\begin{equation}
  E(T) = \der{\ln{\tau_\alpha}}{(1/T)},
\end{equation}
which we compute by the centered difference
method\footnote{Given a set $\{x_i\}$ of points and corresponding function values $\{f_i\}$, with $0\le i\le M$, we compute the derivative at $(x[i+1]+x[i-1])/2$  as $(f[i+1] -
f[i-1])/(x[i+1]-x[i-1])$ if $0<i<M$. We use 
$(f[1] - f[0])/(x[1]-x[0])$ and $(f[M] - f[M-1])/(x[M]-x[M-1])$ 
for the first and last pairs of points, respectively. The expression used at the boundaries is more noisy than the one for $0<i<M$.}.  A graph of $E(T)$
versus $1/T$ provides a simple way to test the parabolic law proposed by
Elmatald et al.~\cite{elmatad_corresponding_2009} in the context of
dynamic facilitation
\begin{equation}
  \label{eqn:par}
  \tau_\alpha = \tau_0 \exp{\left[\left(\frac{J}{T_0}\right)^2
    \left(\frac{T_0}{T}-1\right)^2\right]},
\end{equation}
where $\tau_0$, $J$, and $T_0$ are material parameters.
The activation energy is then a linear function of $1/T$ given by the following expression
\begin{equation}\label{eqn:paraE}
  E_p(T) = \frac{2J^2}{T_0} \left(T_0/T-1\right) .
\end{equation}
It should be emphasized that Eq.~\eqref{eqn:par} is only expected to hold
below the onset temperature $T_0$ and above an additional reference
temperature $T_x$. Outside this range of temperatures, the dynamics is
expected to be Arrhenius~\cite{elmatad_corresponding_2009}.

A similar approach can be used to linearize the classic VFT equation.
Following Stickel et al.~\cite{stickel_dynamics_1995} we introduce
$\phi=E(T)^{-1/2}$.  By computing the derivative of the VFT expression
\begin{equation}
  \label{eqn:vft}
  \tau_\alpha = \tau_0 \exp{\left[ \frac{T_\textrm{vft}}{K(T-T_\textrm{vft})}\right]},
\end{equation}
we obtain
\begin{equation}
  \phi(T) = \sqrt{\frac{K}{T_\textrm{vft}}} \left(1-\frac{T_\textrm{vft}}{T}\right),
\end{equation}
where $\tau_0$, $K$, and $T_{\rm vft}$ are parameters.
Thus, the VFT equation appears linear in a graph where $\phi$ is shown as a function of $1/T$~\cite{stickel_dynamics_1995}.

In Fig.~\ref{fig:tau}(b) we show $E(T)$ as a function of $1/T$.
Overall, the activation energy displays an approximately linear behavior at sufficiently low temperature, i.e., $T<0.5$.
This behavior is well captured by Eq.~\eqref{eqn:paraE}, although  fitting the relaxation times in this range
requires a value $T_0\approx 0.72$, which is lower than the usual estimates
$0.8-1.0$, see e.g. Ref.~\onlinecite{sastry_signatures_1998}, but consistent with a very recent analysis~\cite{Hudson_Mandadapu_2018} based on Eq.~\eqref{eqn:paraE}. 
We also note that the value of $J/T_0=3.7$ is larger than the one used in Ref.~\onlinecite{elmatad_corresponding_2009}. The fit parameters used in Ref.~\onlinecite{elmatad_corresponding_2009} appear to strike a balance between the high and low $T$ portions of the data.
For the most accurate data set, i.e., MD with $N=1200$, and at the lowest temperatures, this representation also reveals a slight inflection in $E(T)$.
This slight saturation may suggest a dynamic crossover similar to the one
observed in certain molecular glass-formers~\cite{novikov_universality_2003} and in the simulation of
harmonic spheres~\cite{kob_nonlinear_2014}.
One comment on possible finite size effects is in order. In a small sample, the activation energy necessarily reaches an upper bound and a smooth crossover to Arrhenius behavior is expected at a system-size dependent temperature~\cite{kim2000apparent}.
Within the quality of our data the temperature dependence for different system sizes gives compatibles results.
However, due to the larger scatter in the $N=3600$ samples, we cannot completely rule out the possibility that the inflection of the $N=1200$ is due to a finite size effect.

\begin{figure}[!t]
  \centering
  \includegraphics[width=\linewidth]{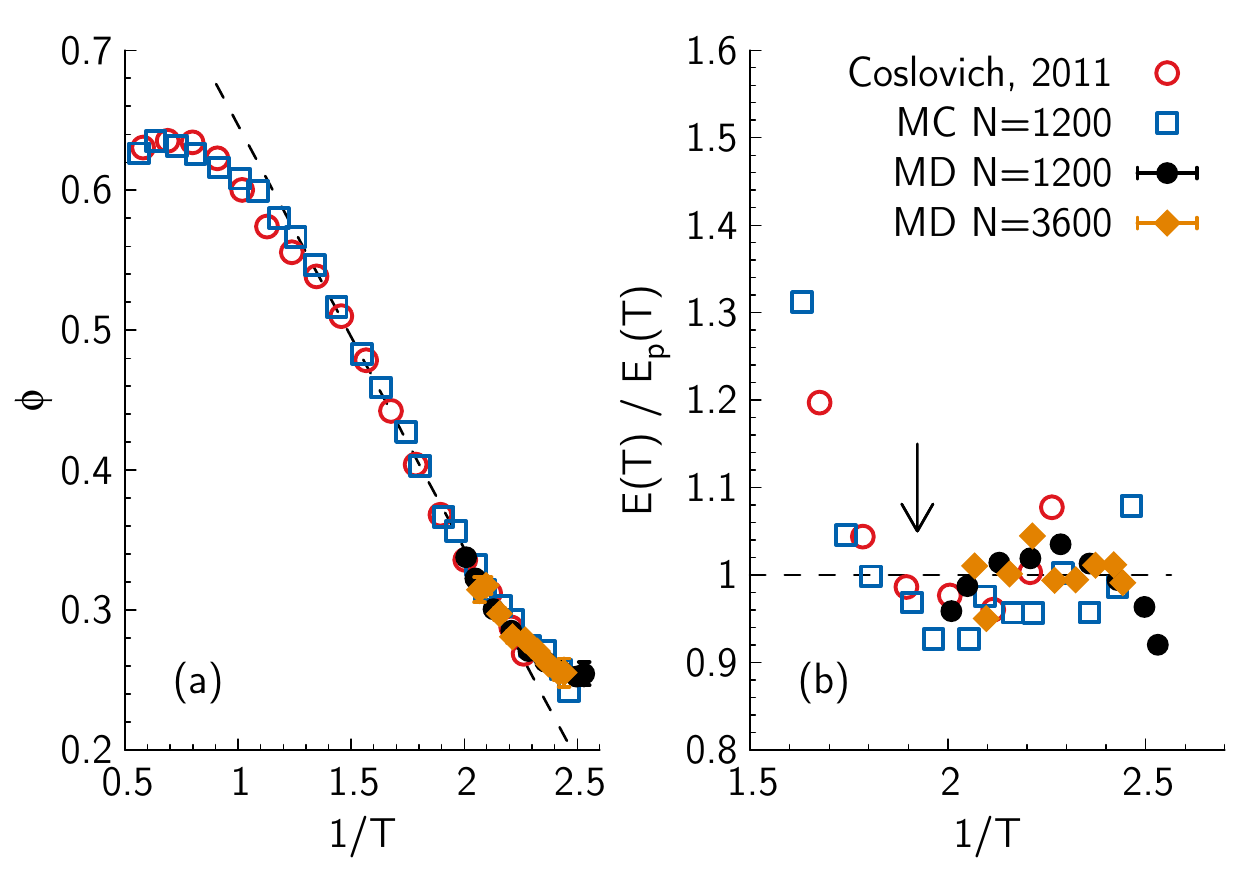}
  \caption{\label{fig:add}Critical assessment of the VFT and parabolic laws. (a) Stickel plot $\phi(T)$ versus $1/T$. (b) Plot of $E(T)/E_p(T)$ as a function of  $1/T$. The dashed lines are guides to the eye. The arrow indicates the temperature $T=0.52$ from which the parabolic law begins to apply.}
\end{figure}

Pushing our analysis one
step further, we compute the derivative of $E(T)$ with respect to $1/T$ and show the results in Fig.~\ref{fig:tau}(c).
This quantity should be constant and equal to $2J^2$ in the dynamic facilitation scenario,
whereas other models~\cite{novikov_universality_2003} predict that it should peak around some
crossover temperature. The inset of the figure illustrates schematically these two possible scenarios.
Within the noise of the data, and in particular
of our most accurate data set (MD with $N=1200$), our measurements are compatible
with a decrease of $d E(T)/d (1/T)$ below some temperature $T_{\rm D}$
close to $T_{\rm MCT}$. Simulations at even lower
temperature and better statistics for larger system sizes would be needed to fully confirm this behavior. 
% Overall, our analysis does not rule out yet either of the dynamic scenarios mentioned above.

In Fig.~\ref{fig:add}(a) we test the validity of the VFT law by computing $\phi(T)$ as a function of $1/T$.
We find that the VFT law holds well at intermediate temperatures, but
clear deviations are visible at low temperature. This confirms the well-known observation~\cite{stickel_dynamics_1995} 
that, except in rare cases, the VFT equation cannot
describe the full $T$ dependence of the dynamic data.
In Fig.~\ref{fig:add}(b) we show a plot $E(T)/E_p(T)$ vs $1/T$. In this
representation, the parabolic law would correspond to a horizontal line. 
We see that the data flatten out only below $T=0.52$, thus the range over which this law holds in the KA mixture is actually more
limited than previously thought~\cite{elmatad_corresponding_2009}.

\begin{figure}[!t]
  \centering
  \includegraphics[width=\linewidth]{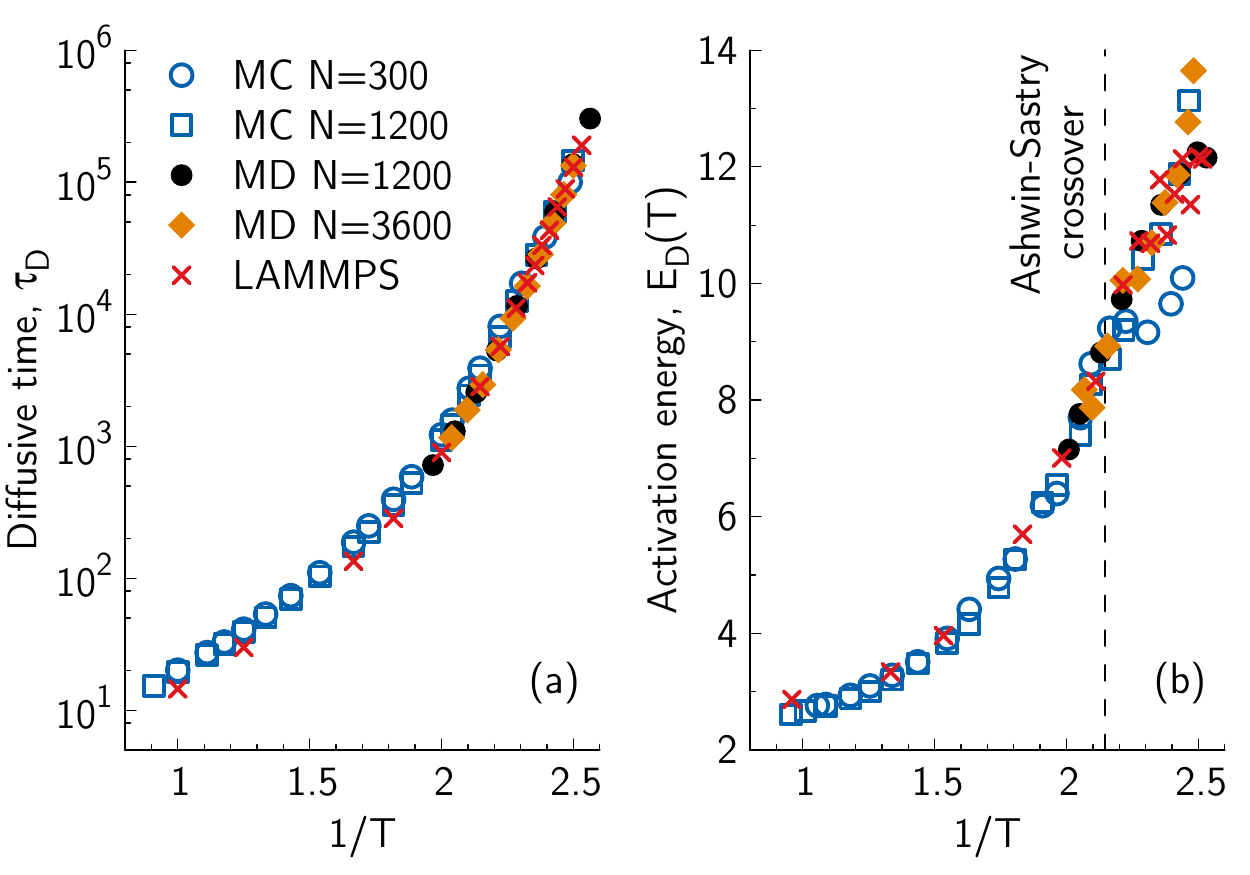}
  \caption{\label{fig:add2}Analysis of diffusive times $\tau_D$. The diffusive times obtained from MC simulations have been scaled by $5\times 10^2$ MC steps. (a) Arrhenius representation of the diffusive times $\tau_D$. (b) Activation energies from the diffusive time $\tau_D$. The vertical line indicates the crossover observed by Ashwin and Sastry in Ref.~\onlinecite{ashwin_low-temperature_2003} for a system of $256$ particles.}
\end{figure}

We also computed the activation energy $E_D$ from
the diffusive time $\tau_D$, defined as the time needed to reach a mean
square displacement of A particles equal to $1$. These diffusive times are shown in Fig.~\ref{fig:add2}(a) for several protocols and system sizes.
We then compute, as before, the activation energies $E_D$ associated to diffusive times $\tau_D$.
Figure~\ref{fig:add2}(b) indicates that the growth of $E_D$ gets milder below 
some crossover temperature $T_{\rm D} \simeq 0.45$. Again, our $N=1200$ MD data-set suggests an inflection of $E_D$ around this temperature.
This crossover temperature is remarkably close to the one reported long ago by Ashwin and
Sastry in Ref.~\onlinecite{ashwin_low-temperature_2003} for a smaller system
size ($N=256$). However, in small systems $E_D$ actually saturates at
$T_{\rm D}$, while bigger samples only cross over to a milder temperature
dependence.  These results indicate that, even though the results of
Ref.~\onlinecite{ashwin_low-temperature_2003} were probably affected by
finite size effects, the diffusion mechanism might actually change in a
temperature range close to the MCT transition temperature. 
This is corroborated by the analysis of a much larger sample ($N=100000$) simulated using LAMMPS,
which gives a trend compatible with the ones observed for $N=1200$ and $N=3600$ particles.
These large-scale simulations were performed in the $NVT$ simulations using the Nose-Hoover thermostat but without parallel tempering.

\begin{figure}[!t]
  \centering
  \includegraphics[width=\linewidth]{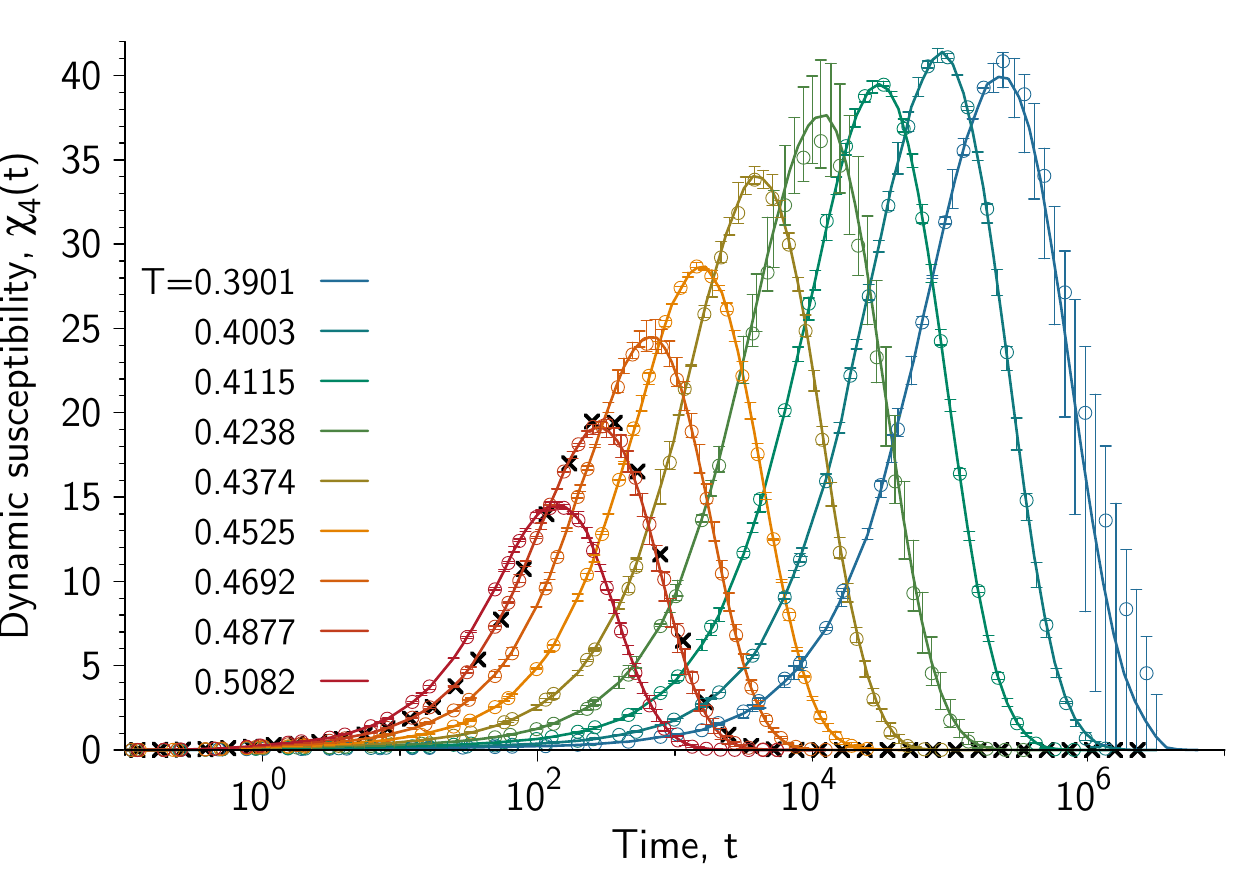} \\
  \caption{\label{fig:chi4t} Dynamic susceptibility $\chi_4(t)$ from MD protcol ($NVT$ ensemble) for $N=1200$ particles. The full lines are measurements over the entire length of the runs, while the symbols are measurements restricted to the first half of the runs. The error bars are estimated from the difference between the first and second halves of the runs. The crosses correspond to $\chi_4^{NVE}(t) + T^2\left(\frac{dQ_s}{dT}(t)\right)^2/c_V$, see Eq.~\eqref{eqn:exact}, at $T=0.4877$.}
\end{figure}

Finally, we compute the dynamic fluctuations associated to the time-dependent self overlap function
\begin{equation}
Q_{\rm s}(t) = N^{-1} \sum_j \Theta(0.3 - |\textbf{r}_j(t)-\textbf{r}_j(0)|) ,
\end{equation}
where the sum is taken over all particles. The dynamic susceptibility is then defined as
\begin{equation}
\chi_4(t) = N \{[\langle Q_s(t)^2 \rangle] - [\langle Q_s(t) \rangle]^2\} ,
\end{equation}
where $[\dots ]$ indicates an average over statistically independent initial samples and $\langle\dots\rangle $ a time average over a given run.
The dynamic susceptibilities, shown in Fig.~\ref{fig:chi4t} for $N=1200$, display a peak at a time $\tau_4$ proportional to the structural relaxation time. The peak height $\chi_4^*$ is a standard proxy for the extent of dynamic heterogeneity in supercooled liquids~\cite{berthier2011theoretical}.

\begin{figure}[!t]
  \centering
  \includegraphics[width=\linewidth]{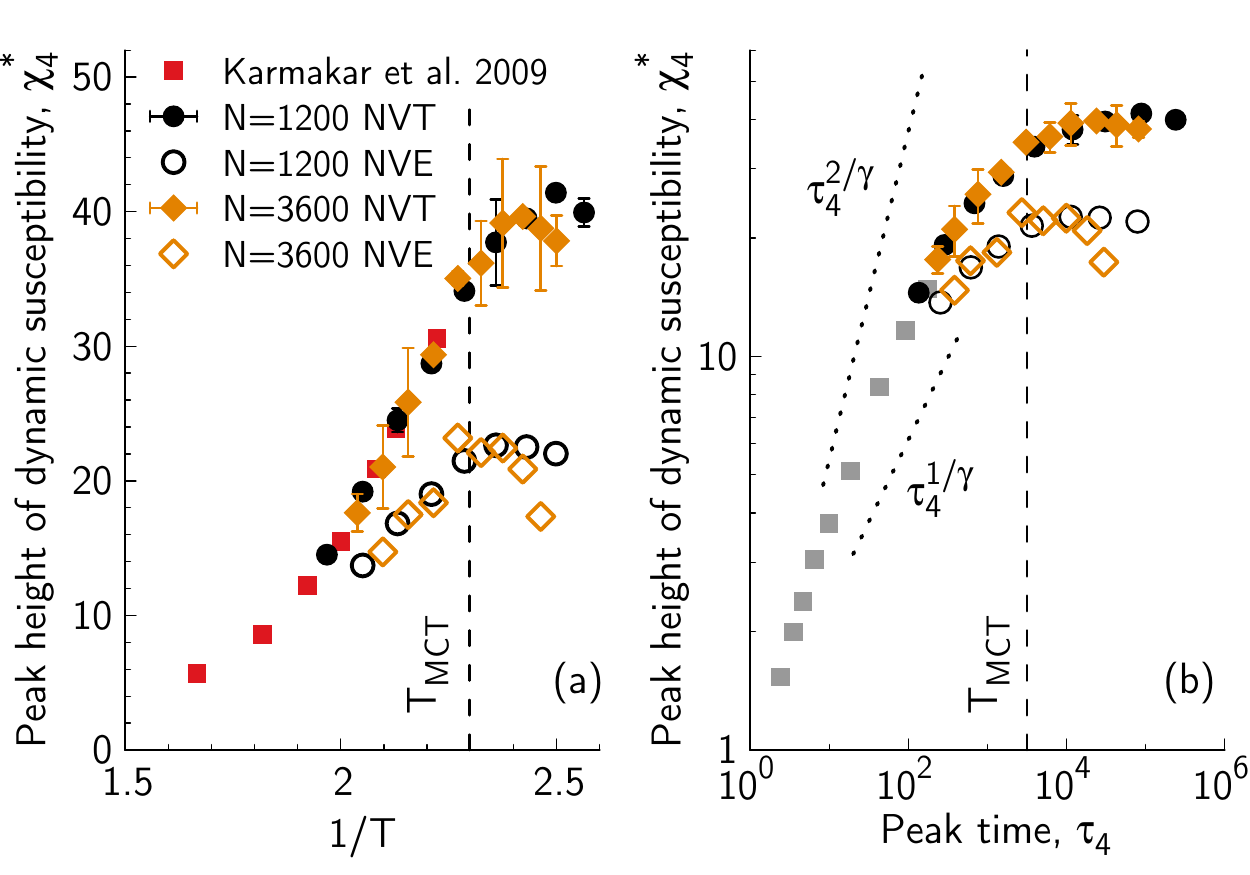} \\
  \caption{\label{fig:chi4}Peak height of the dynamic susceptibility $\chi_4^*$ in the $NVT$ ensemble (full symbols) and in the $NVE$ ensemble (empty symbols) as a function of (a) $1/T$ and (b) the peak time $\tau_4$. The vertical line in the two panels marks the MCT temperature. The dotted lines in panel (b) indicate the predicted MCT scaling in the $NVT$ (exponent $2/\gamma$) and $NVE$ ensemble (exponent $1/\gamma$). Squares in panel (a) are taken from from Ref.~\onlinecite{Karmakar_Dasgupta_Sastry_2009}. Squares in panel (b) indicate results from additional, high temperature $NVT$ simulations for $N=1200$ particles.}
\end{figure}

In Figs.~\ref{fig:chi4}(a) and (b) we show
$\chi_4^*$ as a function of $1/T$ and as a function of $\tau_4^*$, respectively.
The dynamic fluctuations quantified by the susceptibility in the $NVT$ ensemble 
cross over around the MCT temperature to a very mild dependence on both temperature and relaxation time. 
Within the quality of our data, this effect \textit{cannot} be attributed to the finite size of the system.
We point out, however, that the MCT scaling expected in the $NVT$ ensemble~\cite{berthier_spontaneous_2007-2}, $\chi_4^*\sim \tau_{\alpha}^{\gamma/2} \sim \tau_4^{\gamma/2}$, where $\gamma\approx 2.4$ is the exponent of power law fit to the relaxation time, does not hold well for this system. Therefore, the fact that the crossover in the dynamic susceptibility occurs close to the mode-coupling temperature may be coincidental.
We also note that similar results may not hold for the full dynamic susceptibility~\cite{berthier2011theoretical}, which includes contributions from number and concentration fluctuations and which we have neglected here. Our results resonate with those of
Flenner and Szamel~\cite{flenner_dynamic_2013}, who showed that
some contributions to the total susceptibility, notably those associated to dynamic fluctuations in the $NVE$ ensemble,
may be more sensitive to the presence of a crossover than the full susceptibility. 
To address this point quantitatively, we computed the dynamic susceptibility in the $NVE$ ensemble from the exact expression~\cite{berthier_spontaneous_2007-1}
\begin{equation}\label{eqn:exact}
  \chi_4^{NVE}(t) = \chi_4^{NVT}(t) - \frac{T^2}{c_V}\left(\frac{dQ_s}{dT}(t)\right)^2 .
\end{equation}
We checked that the above expression yields results consistent with those obtained from simulations in the $NVE$ ensemble at $T=0.4877$, at which the energy drift during $NVE$ simulations was not too severe, see Fig.~\ref{fig:chi4t}.
The temperature dependence of the peak height of $\chi_4^{NVE}(t)$, shown in Fig.~\ref{fig:chi4}, provides clear hints of a dynamic crossover around the MCT temperature and corroborates the findings of Ref.~\onlinecite{flenner_dynamic_2013}.
Royall et al.~\cite{royall_strong_2015} have also noted that the growth of $\chi_4^*$
observed in simulations at above $T_{\rm MCT}$ is incompatible with the
typical correlation lengths measured experimentally. From this point of view, our finding of a crossover to a
much milder rate of growth of dynamic correlations is a welcome result, which may solve
the apparent conundrum evidenced in Ref.~\onlinecite{royall_strong_2015}.

\section{Conclusions}
\label{sec:conclusions}

In this work we have performed extensive computer simulations to probe the existence of thermodynamic and dynamic anomalies in the Kob-Andersen binary Lennard-Jones mixture. Our simulations build on an efficient simulation setup, which combines algorithmic and hardware optimizations. In particular, we exploit the parallel tempering algorithm and multi-GPU acceleration to measure the specific heat at equilibrium down to unprecedented temperatures. We also appreciably extend the range of dynamic measurements by starting regular MD and MC simulations from independent configurations previously equilibrated with parallel tempering. This approach is trivially parallel and enables us to perform a temperature-derivative analysis of the dynamic data.

Thanks to these advances, we could clarify some issues related to the thermodynamic and dynamic behavior of the mixture. In particular, we found that the specific heat increases monotonically down to at least $T=0.38$, in contrast with previous findings~\cite{flenner_hybrid_2006}, which suggested a maximum around $T=0.43$ and which was probably due to out-of-equilibrium effects. Although crystallization renders the analysis difficult at low temperature, our data indicate that thermodynamic anomalies, if present at all, must occur in a highly metastable, low-temperature portion of the phase diagram.
Conversely, any possible crossover above $T=0.38$ must be purely dynamic in origin.
By performing a temperature-derivative analysis of the dynamic data and by analyzing the dynamic susceptibilities, we have assessed several scenarios that may involve a dynamic crossover in this temperature regime.
One possible scenario suggests the presence of a crossover~\cite{berthier_finite-size_2012,rizzo2016dynamical} around the mode-coupling temperature $T_\textrm{MCT}$. There are however alternative interpretations that attribute
the dynamic crossover to a saturation of energy barriers, as is the case in elastic
models~\cite{dyre_colloquium:_2006,mirigian2014elastically}, free-volume
models~\cite{cohen_liquid-glass_1979} and even possibly in the dynamic facilitation scenario at
sufficiently low temperature~\cite{elmatad_corresponding_2010}. In these latter scenarios, the crossover need not be around $T_\textrm{MCT}$

Although our analysis is not entirely conclusive yet, we argue that state-of-the-art simulation methods combined with temperature-derivative analysis of simulation data may hold the key to fully disentangle these scenarios in the near future. Judging from the quality of our temperature-derivative analysis and dynamic fluctuations measurements, we infer that simulations of the order of several thousands of structural relaxation times are needed to fully confirm or rule out the presence of dynamic crossovers. Our results also suggest that contributions to the dynamic susceptibility measured in ensembles where conserved quantities are not free to fluctuate may better probe the dynamic behavior than the full susceptibility~\cite{berthier_spontaneous_2007-2}. For specific systems, specialized Monte Carlo moves, like particle swaps~\cite{gazzillo1989equation,sindzingre1989calculation,grigera2001fast,ninarello2017models,berthier2016equilibrium}, may provide an additional and important efficiency improvement and facilitate thermodynamic and dynamic studies below the mode-coupling temperature.

\begin{acknowledgments}
We thank E. Flenner and G. Szamel for useful discussions.
We acknowledge PRACE for awarding us access to Curie at GENCI@CEA, France, which enabled the development of part of the software used in this work.
Data relevant to this work can be accessed at
\href{https://doi.org/10.5281/zenodo.1227831}{{https://doi.org/10.5281/zenodo.1227831}}.
\end{acknowledgments}

The final publication is available at Springer via \href{http://dx.doi.org/10.1140/epje/i2018-11671-2}{http://dx.doi.org/10.1140/epje/i2018-11671-2}, albeit with bitmapped figures.

\section*{Author contribution statement}
The simulations were performed as follows: protocol PT-1 and LAMMPS by WK,
protocol PT-2, PT-3, and MD by DC and MC by MO.  DC and MO analyzed
the data. DC wrote the first draft of the article. All authors
discussed the results and contributed to the final manuscript.

\bibliography{ka_lowT}

\end{document}